\documentclass[12pt]{article}
\usepackage{amssymb}
\usepackage{graphicx}
\usepackage{amsmath}

\def\d{\partial}
\def\l{\left(}
\def\r{\right)}

\newcommand{\be}{\begin{equation}}
\newcommand{\ee}{\end{equation}}
\newcommand{\ba}{\begin{align}}
\newcommand{\ea}{\end{align}}
\newcommand{\bg}{\begin{gather}}
\newcommand{\eg}{\end{gather}}
\newcommand{\bseq}{\begin{subequations}}
\newcommand{\eseq}{\end{subequations}}

\renewcommand{\Im}{\mathop{\rm Im}\nolimits}

\def\half{\frac{1}{2}}

\begin{document}

\title{Scalaron the mighty: \\
producing dark matter and baryon asymmetry \\ at reheating }

\author{
D.~S.~Gorbunov\thanks{{\bf e-mail}: gorby@ms2.inr.ac.ru},
A.~G.~Panin\thanks{{\bf e-mail}: panin@ms2.inr.ac.ru}
\\
{\small{\em
Institute for Nuclear Research of the Russian Academy of Sciences,
}}\\
{\small{\em
60th October Anniversary prospect 7a, Moscow 117312, Russia
}}
}
\date{}

\maketitle

\begin{abstract}
In $R^2$-inflation scalaron slow roll is responsible for the
inflationary stage, while its oscillations reheat the Universe. We
find that the same scalaron decays induced by gravity can also
provide the dark matter production and leptogenesis. With $R^2$-term
and three Majorana fermions added to the Standard Model, we arrive at
the phenomenologically complete theory capable of simultaneously
explaining neutrino oscillations, inflation, reheating, dark matter
and baryon asymmetry of the Universe. Besides the seesaw mechanism in
neutrino sector, we use only gravity, which solves all the problems by
exploiting scalaron.
\end{abstract}

\section{Introduction and summary}

The first inflationary model widely discussed in literature, dubbed
$R^2$-inflation \cite{starobinsky}, works in a very economic way,
exploiting one and the same interaction---gravity---to accomplish both
inflation and subsequent reheating. It is tempting to exploit gravity
somewhat further, addressing two other important issues: dark matter
production and generation of the baryon asymmetry of the Universe.

To this end we consider in this work the universal mechanism of particle
production operating in $R^2$-inflationary model: scalaron decay. We
have found that free scalars heavier than 10~keV and fermions heavier
than $10^7$~GeV are forbidden in this scenario, as they would 
overclose the Universe. The fermion of $10^7$~GeV is a viable
dark matter candidate in this model, while light scalars could 
contribute to the {\it hot} dark matter component at best.

We have also found that with two additional right-handed sterile
neutrinos one can explain the neutrino oscillations (via standard
seesaw mechanism) and baryon asymmetry of the Universe (via standard
non-thermal leptogenesis). Curiously, the amount of baryon 
  asymmetry available in the model is strongly constrained from above,
  so that the observed amount is only one order of magnitude below the
  model upper limit. 

Both the dark matter fermions and sterile neutrinos are
produced in post-inflationary Universe in scalaron decays. This
suggests that SM, supplemented with $R^2$-term and three right-handed
sterile neutrinos, forms a kind of {\em naturally complete }
theory. This theory explains different phenomena beyond
the SM---inflation, reheating, dark matter, baryon asymmetry of the
Universe---involving {\em one and the same mechanism } 
based on the peculiarities of 
scalaron interactions with itself and other fields. These phenomena
together with neutrino oscillations are the main observational facts
pointing at incompleteness of the SM. All of them can be explained 
within the proposed model.

\section{Gravitational production of Dark Matter in $R^2$-inflation}

We start with the following Lagrangian in the Jordan frame:
\be
\label{T1*}
S^{JF}=-\frac{M_P^2}{2}\int \!\! \sqrt{-g}\,d^4x\, \l R-\frac{R^2}{6\,\mu^2} \r +
S^{JF}_{matter}\;,
\ee
where we use the reduced Planck mass $M_P$ related to the Planck mass $M_{Pl}$
as $M_P=M_{Pl}/\sqrt{8\pi}=2.4\times 10^{18}$~GeV;
$S^{JF}_{matter}$ includes the action of the Standard Model
and other new fields.
In particular, for free 
scalar $\varphi$ and Dirac fermion $\psi$
one has\footnote{These fields are free in the sense of particle
  physics: in both the Jordan (Eqs.~\eqref{T1**} and \eqref{T1+})   
and Einstein frames (Eqs.~\eqref{T2*} and \eqref{T2+}), 
they are free   at $M_P\to\infty$.}   
\begin{align}
\label{T1**}
S^{JF}_\varphi&=\int \!\! \sqrt{-g}\,d^4x\,
\l
\half \, g^{\mu\nu}\d_\mu\varphi\d_\nu\varphi
-\half \, m_\varphi^2\varphi^2
\r\;,\\
\label{T1+}
S^{JF}_\psi&=\int \!\! \sqrt{-g}\,d^4x\,
\l
i\bar\psi
\hat{\cal D}
\psi
-m_\psi\bar\psi\psi
\r\;,
\end{align}
where $\hat{\cal D}$ is the covariantly generalized Dirac operator,
see e.g.~\cite{Weinberg3}.
It is convenient to go to the Einstein frame by the conformal
transformation
\[
g_{\mu\nu}\to \tilde{g}_{\mu\nu}=\chi\, g_{\mu\nu}\;, ~~~~~~
\chi= {\rm exp}\l \sqrt{2/3}\,\phi/M_P\r\;.
\]
Scalar and fermion fields are rescaled then as
\[
\varphi\to\tilde \varphi = \chi^{-1/2}\,\varphi\;,~~~~~~
\psi\to \tilde\psi = \chi^{-3/4}\,\psi\;,~~~~~~  \hat{\cal D} \to
\hat{\tilde{{\cal D}}} = 
\chi^{-1/2} 
\, \hat{\cal D}\;.
\]
In the Einstein frame the gravity action takes the
Einstein--Hilbert form, but additional scalar degree of freedom $\phi$
emerges and couples to all matter fields. Thus the original action
\eqref{T1*} transforms into
\be
\label{T1++}
S^{EF}=\int \!\!\sqrt{-\tilde g}\,d^4x\,
\left[ -\frac{M_P^2}{2}\,\tilde R + \half\,
\tilde{g}^{\mu\nu} \d_\mu \phi \d_\nu\phi
-\frac{3\,\mu^2 M_P^2}{4}\, \l 1-\frac{1}{\chi\l \phi\r} \r^2
\right] + S^{EF}_{matter}\;,
\ee
where $S^{EF}_{matter}$ includes interactions with the field
$\phi$. In particular, now the actions \eqref{T1**} and \eqref{T1+} read 
\begin{align}
\label{T2*}
S^{EF}_\varphi&=\int \!\! \sqrt{-\tilde g}\,d^4x
\l
\half \, \tilde g^{\mu\nu}\d_\mu\tilde\varphi\d_\nu\tilde \varphi
-\frac{1}{2\,\chi}\, m_\varphi^2\tilde \varphi^2
+\frac{\tilde\varphi^2}{12\,M_P^2}\, \tilde
g^{\mu\nu}\d_\mu\phi\d_\nu\phi
+\frac{\tilde\varphi}{\sqrt{6}\, M_P}\,
\tilde g_{\mu\nu} \d_\mu \tilde\varphi \d_\nu\phi
\r\;,\\
\label{T2+}
S^{EF}_\psi&=\int \!\! \sqrt{-\tilde g}\,d^4x
\l
i\bar{\tilde \psi}
\hat{\tilde{ {\cal D}}}
\tilde\psi
-\frac{m_\psi}{\sqrt{\chi}} \, \bar{\tilde\psi}\tilde\psi
\r\;.
\end{align}
At small values of $\phi$ this new degree of freedom decouples from
all other fields and both frames become identical, as in this limit
$R$ is also small.

Cosmology of the homogeneous and isotropic Universe described by the
action \eqref{T1*} has inflationary
stage \cite{starobinsky} at large values of $R$. In the Einstein frame
this stage is realized as large-field inflation in the slow-roll
regime taking place at 
super-Planckian values of the field $\phi$ which serves as  the inflaton. The
equivalent scalar mode in the action \eqref{T1*}
was named scalaron \cite{starobinsky} and we use both names in what
follows.

Inflation gives rise to primordial scalar perturbations,
whose amplitude normalization to the observed CMB anisotropy and
large-scale structure yields the estimate~\cite{Faulkner:2006ub}
\[
\mu= 1.3\times 10^{-5}\; M_P\;.
\]
The spectral index of the scalar perturbations and
parameters of the generated tensor perturbations are
 consistent with
observational constraints~\cite{Komatsu:2008hk}.

When the slow-roll conditions get violated, inflation terminates
and the inflaton $\phi$ starts to oscillate rapidly, with frequency
equal to scalaron mass $\mu$. This drives the
Universe expansion like at matter-dominated stage. 
The 
intermediate stage naturally ends up with inflaton decays into
ordinary particles due to {\em universal } interactions as in 
\eqref{T2*}, \eqref{T2+}. In particular, scalaron decay rates into a 
pair of sufficiently light scalars and into a pair of sufficiently
light fermions are\footnote{These estimates are in agreement 
with similar ones in the Jordan frame originally obtained in 
\cite{starobinsky,Vilenkin:1985md}, while \eqref{T4+} and \eqref{T4++}
are two times larger and four times
smaller, respectively, than the estimates in Ref.~\cite{Faulkner:2006ub}.} 
\begin{align}
\label{T4+}
\Gamma_{\phi\to \varphi\varphi}&=\frac{\mu^3}{192\pi\,M_P^2} 
\;,\\
\label{T4++}
\Gamma_{\phi\to\bar\psi\psi}&=\frac{\mu\,m_\psi^2}{48\pi\,M_P^2}\;,
\end{align}
respectively. 
In this {\em universal } way the energy flows from the inflaton to
ordinary particles. There is no any amplification of the energy drain to
the SM particles due to coherent effects (like ones in
Ref.~\cite{Kofman:1994rk}), since the produced particles 
interact strongly enough.

Formulas \eqref{T4+}, \eqref{T4++} generally mean that scalar
particles 
play the major role in the scalaron decay process.  The dominant contribution
to the decay rate comes from coupling to kinetic term in
Eq.~\eqref{T2*}. There is no similar term in the fermion case, see
Eq.~\eqref{T2+}, because of the conformal invariance\footnote{This
  implies quite a similar situation for a scalar {\it conformally }
  coupled to gravity in the Jordan frame. We do not consider this
  case, but note that the coherent processes at preheating stage we
  alluded to above may be relevant then. }  of the fermion kinetic term.  The
energy transfers to the relativistic scalars by the time
$t_{reh}\gtrsim 1/\sum_s\Gamma_{\phi\to \varphi\varphi}$. We
define the reheating temperature $T_{reh}$ as an effective temperature
of produced relativistic matter at the time of equality between
inflaton and matter energy densities. Then in the
absence of coherent effects mentioned above,
for $N_s$ scalar components
contributing to the inflaton decay the effective
temperature of produced relativistic matter is
\be
\label{T5+}
T_{reh}\approx 4.5\times 10^{-2}\times g_*^{-1/4}\cdot
\l \frac{N_s\,\mu^3}{M_P}\r^{1/2}\;,
\ee
where $g_*$ denotes the number of relativistic species.
In case of the Standard Model with $g_*=106.75$ and $N_s=4$, 
Eq.~\eqref{T5+} gives numerically
\be
\label{T5*}
T_{reh}\approx 3.1\times 10^9~{\rm GeV}\;,
\ee
which is well in the region where the gauge interactions of the
Standard Model are in equilibrium. So, the value 
\eqref{T5*} is the {\em maximum } temperature of the primordial plasma
or the {\em reheating } temperature, indeed.

With low reheating temperature \eqref{T5*} the post-inflationary
matter-dominated stage lasts long enough, so that the subhorizon scalar
perturbations growing at this stage become nonlinear (for a recent
study see e.g. \cite{Jedamzik:2010dq}). One expects that scalarons 
then start to form self-gravitating clumps of linear sizes much smaller than
horizon, so the Universe still expands as at matter domination. The
scalaron overdensity in clumps can be estimated similar to analysis of
dark matter halos and clumps in the 
late-time Universe, see e.g. \cite{LB}. The scalaron density in clumps
is not high enough to initiate scalaron scatterings due to
self-interaction. Therefore, relativistic (SM) particles production can
be described as decays of (non-relativistic) scalarons and the estimate of
reheating temperature \eqref{T5*} remains intact.

The decay rate formulas \eqref{T4+}, \eqref{T4++} show that the
scalaron produces ordinary particles in a universal way, so that
all scalars and fermions of the theory will eventually populate the
expanding Universe. With all non-gravitational
interactions between ordinary particles switched off, the resulting 
abundances would be determined mostly by spin and, for fermions,
by mass of the particles.

Now let us consider a new field which is either free in the Jordan
frame or couples to other fields very weakly, so that these new
particles never equilibrate in the primordial plasma\footnote{ In
  particular, the Planck-scale suppressed nonrenormalizable couplings
  are allowed.}. If stable at cosmological time scale, this particle
is a good candidate to be dark matter.

Given the discussion above, the particle mass is the only free
parameter and hence its value is fixed by the requirement of
comprising all dark matter whose relative contribution to the present
energy density $\rho_c$ is~\cite{Komatsu:2008hk} $\Omega_{DM}\approx 0.223$.
The mass of dark matter particles $m_{DM}$ and their number density at
present $n_{DM,0}$ are related to  $\rho_c$ and $\Omega_{DM}$ as follows
\be
\label{T6+}
m_{DM}=\frac{\Omega_{DM}\,\rho_c}{s_0}\,\frac{s_0}{n_{DM,0}}\;,
\ee
where we introduced the present entropy density $s_0$.

Let us estimate the entropy-to-dark-matter ratio $s/n_{DM}$
by the time of reheating. The dark matter production after inflation
can be described as the decay of non-relativistic scalarons of mass
$m_\phi$ whose number density $n_\phi$ evolves with scale factor
$a=a\l t\r$ at $t\gtrsim t_{reh}$ as
\[
n_\phi\l a\r= \frac{\mu}{2}\,\phi_{reh}^2\,\l \frac{a_{reh}}{a}\r^3\;.
\]
Here $\phi_{reh}$, $a_{reh}$ refer to the values of inflaton field and
scale factor at reheating. We again neglect any coherent effects
related to the new fields and also treat this new decay channel of
scalaron as subdominant one, so the reheating temperature is still given by
Eq.~\eqref{T5*}. Both points are discussed below in due course: we
will see that the
obtained results justify our choice.
Assuming two-body decays of the inflaton to
dark matter particles with decay rate $\Gamma_{\phi\to DM}$ one
writes down the Boltzmann equation for the dark matter density, 
\[
\frac{d}{dt}\,\l n_{DM}\,a^3\r = 2\, n_\phi\!\l a\r \, \Gamma_{\phi\to
  DM} \, a^3\;.
\]
This equation has a solution
\[
n_{DM}\l t_{reh}\r =\frac{\rho_{reh}}{\mu}\, \Gamma_{\phi\to DM} \, t_{reh}\;,
\]
where $\rho_{reh}$ is the total energy density at the time of reheating,
  when the corresponding contributions of inflaton and relativistic
species coincide. The energy density is related to the Hubble parameter at
reheating $H_{reh}$ by 
\[
\rho_{reh}=2\, g_*\,\frac{\pi^2}{30}\, T^4_{reh}=3\, M_P^2 \,
H_{reh}^2 \;. 
\]
Taking for the numerical estimate of the reheating
time
\[
t_{reh}\simeq \frac{1}{\sqrt{3}\,H_{reh}}\;,
\]
we finally obtain for the entropy-to-dark-matter ratio at reheating
\be
\label{T8+}
\frac{s}{n_{DM}}\l T_{reh} \r =
\frac{2\pi\,\sqrt{g_*} }{3\,\sqrt{15}} \,
\frac{\mu}{\Gamma_{\phi\to DM}}\, \frac{T_{reh}}{M_P}\;.
\ee
Since this ratio remains intact at the further hot stages, we have the
estimate
\[
\frac{s}{n_{DM}}\simeq\frac{s_0}{n_{DM,0}}\;.
\]
Eqs.~\eqref{T8+}, \eqref{T6+} imply the following equation for the mass
of dark matter particles universally produced by the scalaron decay,
\be
\label{T8*}
m_{DM}=\frac{\Omega_{DM}\,\rho_c}{s_0}\,
\frac{2\pi\,\sqrt{g_*} }{3\,\sqrt{15}} \,
\frac{\mu}{\Gamma_{\phi\to DM}}\, \frac{T_{reh}}{M_P}\;.
\ee
Then in the cases of scalar \eqref{T4+} and fermion \eqref{T4++} one
obtains the numerical estimates
 \begin{align}
\label{T9+}
m_\varphi&\approx 6.9\;{\rm keV}\!\times\!\l\! \frac{1.3\!\times\!
  10^{-5}\,M_P}{\mu}\!\r^{\!\!1/2} \l\! \frac{N_s}{4}\!\r^{\!\!1/2}
\l \frac{g_*}{106.75} \r^{\!\!1/4} \l\! \frac{\Omega_{DM}}{0.223}\!\r
\l\! \frac{\rho_c}{0.52\!\times\! 10^{-5}\,{\rm GeV}/{\rm
    cm}^3}\!\r, \\
\label{T9*}
m_\psi&\approx1.2\!\times\!10^7\;{\rm GeV}\!\times\! \l \!\frac{\mu}{1.3\!\times\!
  10^{-5}\,M_P}\!\r^{\!\!1/2} \l \!\frac{N_s}{4}\!\r^{\!\!1/6}
\l \frac{g_*}{106.75} \r^{\!\!1/12} \l \!\frac{\Omega_{DM}}{0.223}\!\r^{\!\!1/3}
\l \!\frac{\rho_c}{0.52\!\times\! 10^{-5}\,{\rm GeV}/{\rm
    cm}^3}\!\r^{\!\!1/3}.
\end{align}
Here we have corrected numerical coefficients (by tens of percent)
to match the results obtained
with numerical integration of the Boltzmann equations which consistently
describe both the energy transfer from scalaron to all relativistic
scalars and the dark matter production.

The dark matter particles are produced highly relativistic with 3-momenta
$\sim \mu/2$, which exceed $T_{reh}$ and hence
momenta of particles in the plasma by about four orders of magnitude,
\[
\mu \sim 10^4\times T_{reh}\;.
\]
So, the free scalars of 10\,keV mass would contribute to {\em hot dark
  matter } component only. This component is certainly subdominant,
hence free scalars cannot solve the dark matter problem in the case of
$R^2$-inflation. The viable choice of dark matter is free fermions of
mass $m_{DM}\approx 10^7$\,GeV. These particles would form {\em cold
  dark matter. }

We end up the discussion of universal dark matter in
$R^2$-inflationary model with several remarks. First,
for the considered case of free boson particles coherent
effects \cite{Kofman:1994rk} during scalaron oscillations could change
the obtained results. However, it only makes worse the situation with
scalar dark matter, and we do not consider these processes.

Second, if at a later stage of the Universe expansion
the entropy gets produced (say, due to decays of nonrelativistic
particles out of equilibrium, etc.), the estimates \eqref{T9+},
\eqref{T9*} have to be corrected to account for the corresponding dilution
factor $r_s=s_{new}/s_{old}$. Thus, r.h.s. of \eqref{T9+} has to
be divided by $r_s$, while r.h.s. of \eqref{T9*} has to be divided
by $r_s^{1/3}$. In realistic models with moderate (if any) entropy
production this does not save scalar dark matter, and only mildly
changes the prediction for the fermion dark matter mass \eqref{T9*}.

Third, in the extensions of particle physics with additional scalars
coupled to the SM fields, 
both reheating temperature \eqref{T5+} and would-be scalar dark
matter mass increase. With very large number of new scalars 
this makes the scalar dark matter viable.

Fourth, if there are other sources of dark matter in the
model---other stable particles or other mechanisms of out-of-equilibrium
production of heavy fermions---the estimate \eqref{T9*} implies the
{\em upper } limit on the mass of universal dark matter in
$R^2$-inflationary model.

Fifth, in case of the Majorana fermions the mass is larger by a factor of
$2^{1/3}$ as compared to the Dirac case \eqref{T9*}, since the decay
rate of scalaron into the Majorana particles is two times smaller. 

Sixth, the estimates \eqref{T9+}, \eqref{T9*} mean that {\em heavier free
particles are forbidden } in models with $R^2$-inflation, otherwise the
Universe would be overclosed. This conclusion is true for the
particles lighter than the scalaron, when these estimates of particle
production rates are applicable. Clearly, production rate of heavier
particles by means of scalaron oscillations is suppressed as compared
to \eqref{T4+}, \eqref{T4++}. The heavier the particles, the stronger
the suppression. Hence, free particles of mass $m\gg \mu$ are
not forbidden in $R^2$-inflation. Moreover, with strongly (most
probably exponentially) tuned value of mass $m\sim\mu$ they can form
viable dark matter. We do not study this situation here.

\section{Leptogenesis in $R^2$-inflation}

The study presented in the previous Section revealed that heavy 
free (in the Jordan frame) fermion is a viable dark matter candidate
in $R^2$-inflationary model. These fermions are naturally produced in
the post-inflationary Universe by scalaron decays, so the right amount
of dark matter is achieved with the mass of about $10^7$\,GeV
\eqref{T9*}. One can further
ask whether it is possible to make use of this universal production
mechanism to unravel another cosmological problem which SM fails to
solve: the baryon asymmetry of the Universe. The answer is positive,
and we illustrate it with example of nonthermal leptogenesis via
decays of heavy sterile 
neutrinos~\cite{Fukugita:1986hr} universally produced
by scalaron decays. This particular example is strongly motivated by
observed oscillations of active neutrinos, phenomena which also
lack an explanation within SM. Heavy sterile neutrinos give masses to
active neutrinos via seesaw mechanism. This mechanism is responsible for the
hierarchy between the neutrino masses and electroweak scale. In this
Section we show that both the correct neutrino masses and successful 
leptogenesis can be realized 
in the $R^2$-inflationary model.

The modification of SM we consider consists of two new Majorana
fermions $N_I$, $I=1,2$, which are right singlets with respect
to the SM gauge group. The most general renormalizable lagrangian for
these fermions is
\be
\label{c2+}
{\cal L}= i \bar N_I \gamma^\mu\d_\mu N_I - y_{\alpha I} \bar L_\alpha
N_I \tilde \Phi - \frac{M_I}{2}\bar N_I^cN_I + h.c.\;,
\ee
where $y_{\alpha I}$ are new Yukawa couplings, $\Phi$ is the SM Higgs
doublet and $\tilde \Phi=\epsilon\,\Phi^*$ with $\epsilon$ being
antisymmetric $2\times2$ matrix.

When electroweak symmetry breaks down, $\Phi$ acquires vacuum
expectation value, $\Phi^T=\l 0, v/\sqrt{2}\r$, $v=246$~GeV. Then
\eqref{c2+} yields active-sterile mixing in the neutrino mass
matrix. Assuming $yv\ll M_I$, this mixing gives for the active
neutrino masses:
\be
\label{c2*}
{m_{\nu}}_{\,\alpha\beta}= - \sum_Iy_{\alpha I}\frac{v^2}{2\,M_I}y_{\beta I}\;.
\ee
Hence, $m_\nu\ll v$, that is the seesaw 
mechanism\footnote{Note that with only two seesaw sterile neutrinos we
end up with one out of three active neutrino being massless.}. 
The formula
\eqref{c2*} together with atmospheric neutrino mass
splitting, $\Delta m_{atm}^2\simeq 3\times
10^{-3}$~eV$^2$~\cite{Amsler:2008zzb}, implies an order-of-magnitude
upper limit on the {\em lightest sterile neutrino } mass:
\be
\label{c3*}
M_{lightest}\lesssim \frac{v^2}{2\sqrt{\Delta m_{atm}^2}}\simeq 5\times
  10^{14}~{\rm GeV}\;.
\ee
The smaller the Yukawa couplings $y_{\alpha I}$ and the heavier the
active neutrinos, the smaller $M_{lightest}$, see Eq.~\eqref{c2*}.

Lagrangian \eqref{c2+} added to SM explicitly violates the lepton
number. The source of CP-violation, which is one of
the Sakharov's conditions~\cite{Sakharov:1967dj} 
for successful baryo- and leptogenesis, is the complex Yukawa
couplings $y_{\alpha I}$. The third Sakharov's condition, departure
from thermal equilibrium for lepton number violating processes, is
achieved due to decay of non-relativistic sterile
neutrinos, which is evidently an out-of-equilibrium process.

Let us discuss the production of these heavy sterile neutrinos in the
post-inflationary Universe with oscillating scalaron. Neutrinos are
produced via Yukawa-type coupling to scalaron, see
Eq.~\eqref{T2+}. The production is effective only for neutrinos
lighter than scalaron. Indeed, the amplitude of scalaron oscillation
never exceeds $M_P$. At the same time the scalaron coupling to a
fermion is proportional to the fermion mass, so the latter remains
almost intact. Therefore, an enhancement of heavy fermion production
by Yukawa source due to periodic decrease of fermion effective
mass~\cite{Giudice:1999fb} does not work here.

Thus, we consider the models with sterile neutrinos lighter than
scalaron, $M_I<m_\phi$. By the time of reheating, $t\simeq t_{reh}$,
the ratio of number density of produced in scalaron decays heavy
neutrinos to entropy is given by Eq.~\eqref{T8+}. Substituting
relevant numbers in \eqref{T8+} and \eqref{T4++} (recall that for
Majorana fermions the decay rate is two times lower than for the Dirac
fermions) one obtains
\be
\label{c3+}
\frac{n_{N_I}}{s}\l T_{reh}\r=2.9\times 10^{-6}\times \l
\frac{M_I}{5\times 10^{12}~{\rm GeV}}\r^2\;, 
\ee
where we have included a correction of about tens of percent in order to match
the solution of the relevant Boltzmann equations (see comment below
Eqs.~\eqref{T9+}, \eqref{T9*}).

These sterile neutrinos, however, decay quite rapidly due to the same 
Yukawa couplings in \eqref{c2+} responsible for the seesaw
mechanism. Their total decay rates are
\[
\Gamma_{N_I}=\frac{M_I}{8\pi}\sum_\alpha \left| y_{\alpha I} \right|^2 \;.
\] 
Disregarding any hierarchy in the neutrino Yukawa sector one takes
for the order-of-magnitude numerical estimate of Yukawa
couplings relation \eqref{c2*} with $m^2_\nu\sim \Delta m^2_{atm}$
and finds for the lightest sterile
seesaw neutrino
\[
\Gamma_{N_I}\sim \frac{\sqrt{\Delta m^2_{atm}}}{4\pi}
\frac{M_I^2}{v^2}\;.
\]
Thus, sterile neutrino of mass above $10^{10}$~GeV decays before
reheating\footnote{Eqs.~\eqref{c3+} and \eqref{T5*} mean that sterile
  neutrino decays contribute not much to the Universe reheating.}. 
This is a very process which generates lepton asymmetry in
the early Universe. To simplify formulas we assume further, that only
the lightest neutrino, $N_1$, contributes and  $M_1\ll M_2$. Then
the value of lepton asymmetry $\Delta_L=n_L/s$ is
\be
\label{c4+}
\Delta_L= \delta_L\cdot \frac{n_{N_1}}{s}\;,
\ee
with~\cite{Fukugita:1986hr} (for reviews see
e.g. \cite{Buchmuller:2005eh}) 
\be
\label{c4*}
\delta_L=-\frac{3\,M_1}{8\,\pi v^2 } \cdot
\frac{1}{\sum_{\alpha,\, I} \left| y_{\alpha I}\right|^2}
\sum_{\alpha\beta} \Im \l y_{\alpha 1} y_{\beta 1} y_{\alpha 2}^*
\, \frac{v^2}{2M_2}\,y_{\beta 2}^*  \r
\sim \frac{3M_1\,\sqrt{\Delta m^2_{atm}}}{8\pi v^2}\;,
\ee
where we have used the order-of-magnitude estimate \eqref{c2*}
with $m^2_\nu\sim \Delta m^2_{atm}$. Finally,
from \eqref{c4+}, \eqref{c4*} and \eqref{c3+} we obtain
\be
\label{c4++}
\Delta_L\sim 1.5\times 10^{-9}\times
\l \frac{M_1}{5\times 10^{12}~{\rm GeV}}\r^3\;.
\ee

The asymmetry is generated mostly at late times just before reheating,
when the ratio of sterile neutrino (if stable) density to entropy 
becomes maximum, see Eq.~\eqref{T8+}.
The sterile neutrinos are quite heavy so the lepton
asymmetry is not washed out by inverse decay processes, which are out
of equilibrium. Likewise, another
important lepton violating process, scatterings $\Phi L
\leftrightarrow \Phi \bar L$ through virtual sterile neutrinos (with
cross section $\sigma \sim \frac{m_\nu^2}{8\pi\, v^4}$)  
are out-of-equilibrium at the epoch of interest. Thus, lepton
asymmetry \eqref{c4++} transfers later on into baryon asymmetry by sphaleron
processes \cite{Khlebnikov:1996vj},
\[
\Delta_B\simeq\Delta_L/3 \sim 0.5\times 10^{-9}\;.
\]
Given the present baryon asymmetry of the Universe
$\Delta_{B,0}=0.88\times 10^{-10}$~\cite{Komatsu:2008hk} and
the estimates above, we conclude that heavy seesaw
sterile neutrinos of mass $10^{12}$-$10^{13}$\,GeV, produced in scalaron
decays, in $R^2$-inflationary model can be responsible for the baryon
asymmetry of the Universe. Since the sterile neutrino are lighter than
scalaron, the available room is not large, that one
could interpret as {\it an order-of-magnitude prediction of the baryon
  asymmetry, } which is otherwise (and usually) appears as an
accidental number.

The estimates above are an order-of-magnitude only. In the  
particular model under discussion with only two seesaw sterile
neutrinos one can relate all the Yukawa couplings to the measured active
neutrino mass differences and mixing angles, so that only few model
parameters remain free. In this way one can refine our estimates.
Likewise the numerical estimates may be changed in models with a particular
hierarchy of Yukawa couplings in the neutrino sector.

Note in passing that second, heavier sterile neutrino $N_2$ if not too
heavy can also
contribute to lepton asymmetry generation, its decay products do not
equilibrate.
There is also some contribution from sterile neutrino decays at
earlier times, when they are relativistic. The inverse processes are
rare because the particle density is small.

\section{Discussion}

To conclude, we have shown that in $R^2$-inflation the scalaron decay
responsible for reheating can also take care of dark matter production
and generation of the baryon asymmetry of the Universe. The latter
occurs through nonthermal leptogenesis by introducing seesaw
neutrinos. Thus, neutrino oscillations are also explained in this
model, which pretends to be {\it phenomenologically self-contained. }

We end up with several comments. First, at relatively low energy
scale, 
the model we discussed can be further modified in gravity sector
(attempting to explain the late-time accelerated expansion of the
Universe, as e.g. 
in Refs.~\cite{Nojiri:2003ft,Starobinsky:2007hu}) or in particle
physics sector (say, by adding $PQ$-axion \cite{Peccei:1977hh}
to solve the strong CP-problem). All these and similar modifications can be
safely adopted provided the early time cosmology remains intact.

Second, we do not discuss here the gauge hierarchy problem in particle
physics. Certainly, $R^2$-term does not contribute to this problem:
all couplings of scalaron to other particles are suppressed by the
Planck scale. With the absence of new scales in {\em particle physics,}  one
cannot discard that still unknown complete quantum theory at gravity scale 
(quantum gravity) 
is responsible for the
cancellation of dangerous quantum corrections (if any) to the Higgs boson
mass. In fact, a nontrivial space structure that shows up at Planck scales
could also invalidate the strong CP-problem (see
e.g. \cite{Khlebnikov:1987zg}). Thus, quantum gravity could
solve all problems, indeed. Otherwise, one can think about other
solutions usually considered. In particular, our logic can be adopted
for supersymmetric extensions of the SM and models with axion. In all
these cases free fermion again serves as dark matter candidate, though
it is somewhat heavier than in the case of SM because of larger
number $N_s$ of scalars responsible for reheating and larger 
number of degrees of
freedom $g_*$, see Eq.~\eqref{T9*}.

Third comment, which is related to the second one. 
The heavy sterile neutrinos (of
mass $M_N$) we used contribute to gauge hierarchy problem making it
worse as compared to the SM. Indeed, their coupling to the SM Higgs
boson gives rise to both divergent and finite, of order $M_N$,  
contributions to the Higgs boson mass. This worsening can be avoided if sterile
neutrinos coupled to the Higgs boson have masses of order electroweak
scale or smaller, as e.g. in $\nu$MSM model~\cite{Asaka:2005an}.
Indeed, heavy sterile 
neutrinos can be replaced by the light ones in $R^2$-model we 
discuss. In that case, however, sterile neutrinos, required for
leptogenesis, should be produced not by the scalaron decay (which, as we saw, requires
large mass to be efficient enough), but via neutrino oscillations in
the early Universe~\cite{Akhmedov:1998qx}.
The $R^2$-model with one heavy fermion (free in 
the Jordan frame) and two light seesaw sterile neutrinos can be {\it both
phenomenologically and theoretically self-contained } upto the quantum
gravity effects. At the same time, this modification has also one
more advantage as compared to what we discuss in the main text:
sterile neutrino sector here can be directly
tested~\cite{Gorbunov:2007ak}  in
particle physics experiments.

We thank S. Odintsov, V. Rubakov, M. Shaposhnikov, A. Starobinsky, 
I. Tkachev and M. Vasiliev for
valuable discussions.  The work is supported in part by the grant of
the President of the Russian Federation NS-5525.2010.2 (government
contract 02.740.11.0244).  D.G. thanks ITPP EPFL and the organizers of
the long-term workshop in Yukawa Institute YITP-T-10-01 for
hospitality.  The work of D.G.  is supported in part by the Russian
Foundation for Basic Research (grants 08-02-00473a), by FAE program
(government contract $\Pi$520) and by SCOPES program. The work of
A.P. is supported in part by the "Dynasty" Foundation (awarded by the
Scientific board of ICPFM), by FAE program (government contract
$\Pi$2598) and by the grant of the President of the Russian Federation
MK-4317.2009.2.


\end{document}